\documentstyle[prb,multicol,aps,psfig]{revtex}
\newcommand{\be}{\begin{equation}}
\newcommand{\ee}{\end{equation}}
\newcommand{\calH}{ {\cal H} }
\newcommand{\bfr}{ {\bf r} }

\begin{document}
\title{Spin Polarization of the Low Density 3D Electron Gas}
\author{F. H. Zong, C. Lin,  and D. M. Ceperley\\}
\address{Dept. of Physics and NCSA, University of Illinois at
Urbana-Champaign, Urbana, IL 61801} \maketitle

\begin{abstract}
To determine the state of spin polarization of the 3D electron gas
at very low densities and zero temperature, we calculate the
energy versus spin polarization using Diffusion Quantum Monte
Carlo methods with backflow wavefunctions and twist averaged
boundary conditions. We find a second order phase transition to a
partially polarized phase at $r_s\sim 50 \pm 2$. The magnetic
transition temperature is estimated using an effective mean field
method, the Stoner model.
\end{abstract}

\vspace*{3mm} \noindent PACS Numbers:71.10.Ca, 71.10.Hf, 75.10.Lp,
05.50.Fk

\begin{multicols}{2}
\narrowtext

\section{Introduction}

The three dimensional homogeneous electron gas, also known as the
fermion one component plasma or jellium, is one of the simplest
realistic models in which electron correlation plays an important
role. Despite years of active research, the properties of
thermodynamic phases of the electron gas are still not known at
intermediate densities.\cite{ortiz} In this paper, we study the
spin polarization phase transition of the three dimensional
electron gas at zero temperature with recently improved quantum
Monte Carlo methods.

There has been recent interest in the low density phases spurred
by the observation of a ferromagnetic state in calcium hexaboride
(CaB\( _{6} \)) doped with lanthium\cite{young99}. The magnetic
moment corresponds to roughly 10\% of the doping density. The
temperatures (600K) and densities ($7 \times 10^{19} /cm^3$) of
this transition are in rough agreement with the predicted
transition in the homogeneous electron gas.\cite{ortiz}. However,
to make a detailed comparison, it is necessary to correct for band
effects. For example, conduction electrons are located at the
X-point of the cubic band structure and thus have a six-fold
degeneracy. The effective mass of electrons at this point and the
dielectric constant are also changed significantly from their
vacuum values.\cite{pickett} These effects cast doubt on the
viability of the electron gas model to explain the observed
phenomena. Excitonic models have been proposed to explain the
ferromagnetism\cite{cab6theory}. Whatever the interpretation of
ferromagnetism in $CaB_6$, the determination of the polarization
energy of the electron gas is important problem because of the
importance of the model.

The ground state properties of the electron gas are entirely
determined by the density parameter $r_{s}= a/a_0$ where $4 \pi
\rho a^3/3=1$ and $a_0$ is the bohr radius, possibly changed from
its vacuum value by band effects. In effective Rydbergs, the
Hamiltonian is :\begin{equation} \label{hamiltonian}
H=-\frac{1}{r_{s}^{2}}\sum ^{N}_{i=1}\nabla
_{i}^{2}+\frac{2}{r_{s}}\sum _{i<j}\frac{1}{\left|
\mathbf{r}_{i}-\mathbf{r}_{j}\right| }+const.
\end{equation}
Note that the kinetic energy scales as \( 1/r^{2}_{s} \) and the
potential energy scales as \( 1/r_{s} \) so that for small \(
r_{s} \) (high electronic density), the kinetic energy dominates,
and the electrons behave like ideal gas; in the limit of large \(
r_{s} \), the potential energy dominates and the electrons
crystallize into a Wigner crystal.\cite{wigner34} There is a first
order freezing transition\cite{cep80} at $r_s\approx 100$.

Considering now the spin degrees of freedom, at small \( r_{s} \),
electrons fill the Fermi sea with equal number of up spin and down
spin electrons to minimize the total kinetic energy and thus the
total energy; the system is in the paramagnetic state. As the
density decreases and before the freezing transition, there is a
possibility that the electrons become partially or totally
polarized (ferromagnetic). The spin polarization is defined as \(
\zeta =\left| N_{\uparrow }-N_{\downarrow }\right| /N \), where \(
N_{\uparrow } \) and \( N_{\downarrow } \) are the number of up
and down spin electrons, respectively and \( N=N_{\uparrow
}+N_{\downarrow } \). For paramagnetic phase \( \zeta =0 \) and
for ferromagnetic phase \( \zeta =1 \).

This polarization transition was suggested by Bloch\cite{bloch29}
who studied the polarized electronic state within the Hartree-Fock
(HF) approximation. He found the ferromagnetic state favored over
paramagnetic state for $ r_{s} > 5.45$, almost within the density
of electrons in metals. However, HF is not accurate for $r_s>0$.

More accurate energies became available with the development of
Monte Carlo methods for many-fermion systems. Ceperley\cite{cep78}
using Variational Monte Carlo with a Slater-Jastrow trial function
determined that the transition between the polarized and
unpolarized phase occured at $r_s= 26 \pm 5$. Using a more
accurate method, diffusion Monte Carlo (DMC),\cite{cep80} it was
estimated that the polarized fluid phase is stable at \(
r_{s}=75\pm 5 \). An extension to this work\cite{acp82} found the
\( \zeta =0.5 \) partially polarized fluid becomes stable at
roughly \( r_{s} \approx 20 \) and the completely polarized state
never stable.

Recently Ortiz et al.\cite{ortiz} applied similar
methods\cite{ortizm} to much larger systems ($ N\leq 1930$) in
order to reduce the finite-size error. They concluded the
transition from the paramagnetic to ferromagnetic transition is a
continuous transition, occurring over the density range of \(
20\pm 5\leq r_{s}\leq 40\pm 5 \), with a fully polarized state at
$r_s \geq 40$.

Due to the very small energy differences between states with
different polarizations, systematic errors greatly affect the QMC
results. Recent progress in the quantum simulation methods makes
it possible to reduce these errors. Kwon et al.\cite{kwon98} found
that a wavefunction incorporating back-flow and three-body (BF-3B)
terms provides a more accurate description: they obtained a
significantly lower variational and fixed-node energy. In another
advance of technique, twist-averaged boundary
conditions\cite{lin01} (TA) have been shown to reduce the
finite-size error by more than an order of magnitude, allowing one
to obtain results close to the thermodynamic limit using results
for small values of $N$. In this paper, we apply these improved
methods to the polarization transition in the three dimensional
electron gas. We first describe the simulation method, and then,
the results.

\section{Methods}

The  most accurate Quantum Monte Carlo (QMC) method\cite{rmp-f} at
zero temperature is Projector or Diffusion Monte Carlo (DMC): one
starts with a trial function and uses $\exp(-t H)$ to project out
the ground state using a branching random walk. Fermi statistics
pose a significant problem for the projection method, since exact
fermion methods such as transient estimate or release-node QMC
suffer an exponential loss of efficiency for large numbers of
particles. For this reason, the fixed-node approximation is
normally used, obtaining the best upper bound to the energy
consistent with an assumed sign of the wave function. The
generalization of the fixed-node method to treat complex-valued
trial functions is known as the fixed-phase
approximation.\cite{fp}

In the simpler, but less accurate Variational Monte Carlo method
(VMC), one assumes an analytic form for a trial function $\Psi_T
(R)$ and samples $|\Psi_T (R)|^2$ using a random walk. An upper
bound to the exact ground state energy is the average of the local
energy $E_L (R) = \Psi_T (R) ^{-1} \calH \Psi_T (R)$ over the
random walk.

The trial wavefunction plays a very important role in these two
methods. With a better trial wavefunction, not only is the
variational energy lower and closer to the exact energy, but also
the variance of the local energy is smaller so that it takes less
computer time to reach desired accuracy level. The trial
wavefunction is also important in fixed-phase DMC because the
solution is assumed to have the same phase as the trial function.
One then solves for the modulus. This implies that the DMC energy
lies above the exact ground state energy by an amount proportional
to the mean squared difference of the phase of the trial function
from the exact phase. As this is the only uncontrolled
approximation, it is important to carefully optimize the assumed
trial wavefunction.

For a homogeneous system, the non-interacting (NI) wavefunction
consists of a Slater determinant of single-electron plane waves
orbitals. To incorporate electron correlation, one multiplies the
NI wavefunction by a pair wavefunction, obtaining the so-called
Slater-Jastrow (SJ) form. To construct a better trial function,
one incorporates backflow and three-body effects.\cite{kwon98} The
particle coordinates appearing in the determinant become
quasi-particle coordinates:\be \mathbf{x}_{i}=\mathbf{r}_{i}+\sum
_{j\neq i}^{N}\eta (r_{ij})(\mathbf{r}_{i}-\mathbf{r}_{j}).\ee The
Slater determinant is then $ D=\det (e^{i\mathbf{k}_{m}\cdot
\mathbf{x}_{n}})$ where \( \eta (r_{ij}) \) is a function to be
optimized. Then the backflow-three body wavefunction (BF-3B)
is:\be \Psi _{T}(R)=D_{\uparrow }D_{\downarrow }\exp \left( -\sum
_{i<j}^{N}\widetilde{u}(r_{ij})-\frac{\lambda _{T}}{2}\sum
_{i=1}^{N}\mathbf{G}_i^2\right) \ee where \be \mathbf{G}_i=\sum
_{j\neq i}^{N}\xi (r_{ji})(\mathbf{r}_{j}-\mathbf{r}_{i})\ee and
\begin{equation} \label{utilde}
\widetilde{u}(r)=u^{RPA}(r)-\lambda _{T}\xi ^{2}(r)r^{2}+\gamma
(r).\end{equation} Here \( D_{\uparrow } \) and \( D_{\downarrow }
\) are determinants for the up spin and down spin electrons, \(
\xi (r) \) is the three-body correlation function, and ${\tilde
u}$ is the Jastrow correlation function. For the electron gas an
accurate analytic form,\cite{gaskell61} $u^{RPA}(r)$, has as low
an energy\cite{cep78} as those with optimized parameters. In the
presence of three-body correlation, the RPA two-body term is
supplemented with an extra Gaussian function \( \gamma (r) \).
Please refer to Kwon et al.\cite{kwon98} for further details
concerning this wavefunction. We used optimized Ewald sums
\cite{natoli} both for the potential and for the correlation
factor so as to have the correct long wavelength behavior.

Though the computational cost for BF-3B wavefunction is somewhat
greater than the simple SJ wavefunction and there is the added
cost of optimization for BF-3B wavefunction, we found accurate
trial wavefunctions crucial to compute the small energy
differences between different polarization states. We optimized
the parameters by minimizing a combination of the energy and the
variance for each density and polarization. Figure (1) shows the
energy vs. polarization at \( r_{s}=50 \) using different trial
functions and simulation methods. The SJ trial function with VMC
has the highest energy for all polarizations and at this level of
accuracy finds the fully polarized phase to be stable, in
agreement with earlier VMC calculations.\cite{cep78} However,
using the best BF-3B trial function, the variational energies are
lowered significantly with the unpolarized energy dropping more
than the polarized case so that the polarized phase is no longer
stable. DMC calculations confirm this result. Note that the DMC
energies determined using the NI phases (or nodes) give energies
lower than the BF-3B variational energies confirming the
importance of accurate DMC calculations. The use of BF-3B
wavefunctions with DMC leads to the lowest ground state energies,
hopefully, very close to the exact energy.

\begin{figure}
\centerline{\psfig{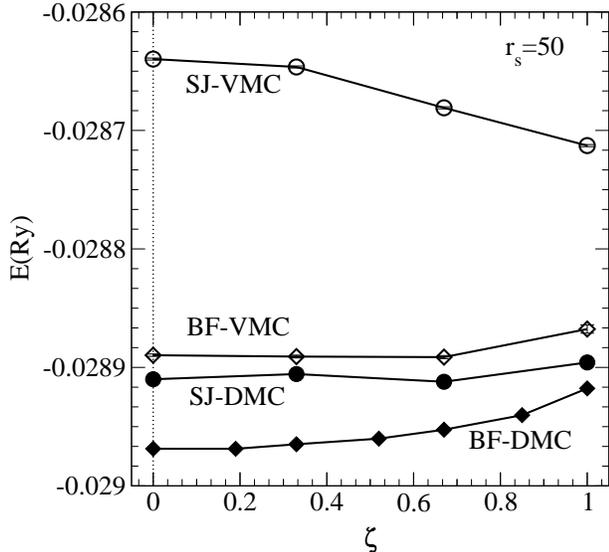}}
%{\centering
%\resizebox*{0.8\columnwidth}{!}{\includegraphics{fig/3deg/rs50/method.eps}}
%\par}
\label{TAmethod} \caption{Energy versus spin polarization at
$r_{s}=50$ for 54 electrons using TA with \( 10^{3} \) twist
values. Compared are calculations with SJ and BF-3B wavefunctions
and with two QMC methods: VMC and DMC. }\end{figure}After the
effect of the nodes, the dependence of the energy on the number of
electrons is the largest systematic error. Within periodic
boundary conditions (PBC), the phase picked up by the wavefunction
as a particle makes a circuit across the unit cell, is arbitrary.
General boundary conditions are: \be \Psi(\bfr_1+L, \bfr_2,
\ldots)= e^{i\theta}\Psi(\bfr_1, \bfr_2, \ldots)\ee where $L$ is a
lattice vector of the supercell. If the twist angle $~\theta$ is
averaged over, most single-particle finite-size effects arising
from shell effects in filling the plane wave orbitals, are
eliminated.  This is particularly advantageous for polarization
calculations since shell effects dominate the polarization energy.
The extra effort in integrating over the twist angles is minimal,
since the various calculations all serve to reduce the final
variance of the computed properties. The effect of boundary
conditions is examined in detail in Lin {\it et al.}\cite{lin01}

\begin{figure}
\centerline{\psfig{figure=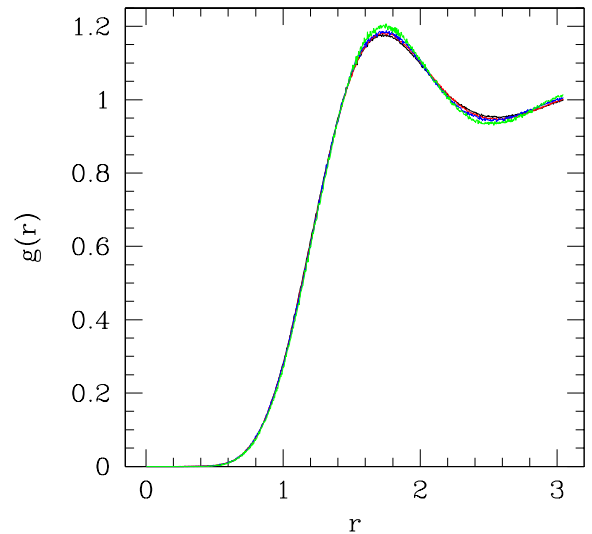,width=10cm,angle=0}}
\caption{\label{gofr}The pair correlation function for several
polarizations at \protect{$r_s=50$} using DMC. The various curves
are for $\zeta = 0, 0.33, 0.67, 1$ with the structure increasing
with spin polarization. }
\end{figure}

\begin{figure}
\centerline{\psfig{figure=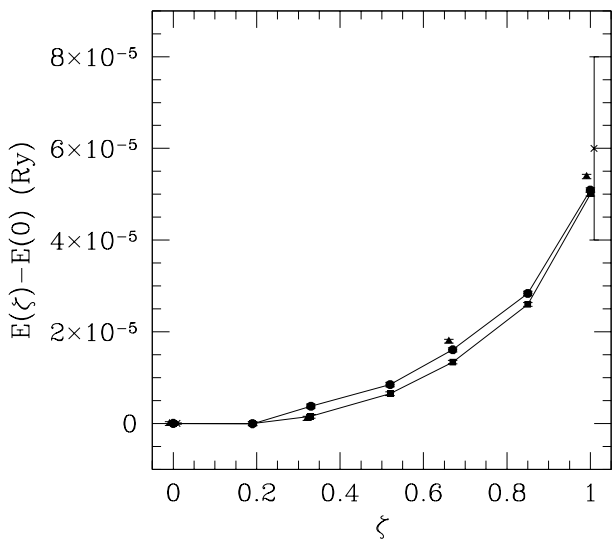,width=10cm,angle=0}}
\caption{\label{TA_50size}The polarization energy for various
sized systems at \protect{$r_s=50$} using TA and  DMC. (circle 54,
square 108, and triangle  162). The point at $\zeta=1$ with the
large error bar is from Ceperley-Alder.\cite{cep80}   Other errors
are less than $10^{-6}$. }
\end{figure}

There is a further size effect in the calculation of the potential
energy due to a charge interacting with its correlation hole in
neighboring supercells as shown in Fig. (6) of Lin {\it et
al.}\cite{lin01}. To correct this, we fit the energies versus $N$
using the expansion:\begin{equation} \label{3deg-TA}
E_{N}=E_{\infty} +\frac{a_1}{N}+\frac{a_2}{N^{2}} + \ldots.
\end{equation}
For unpolarized systems, the fitted $E_{\infty}$ agrees with the
previous non-twist averaged (PBC) result determined using
extrapolations based on fermi liquid theory (FLT). As shown in
Fig. \ref{gofr},  the correlation hole is only weakly dependent on
spin polarization at low density: the peak of $g(r)$ only changes
from 1.175 to 1.190 as the system goes from unpolarized to
polarized. Hence, the potential size-effect hardly changes the
spin polarization energy. Fig. \ref{TA_50size} shows the
polarization energies for $N=54, 108, 162$. With TA boundary
conditions, there is a remarkable insensitivity to the number of
electrons. Even though the system size is increased threefold, the
change in the energy versus polarization is almost undetectable.
Considering only the leading \( 1/N \) correction, we estimated \(
E_{\infty } \) with $N=54$ and $N=108$.

Also shown in Fig. \ref{TA_50size} is the estimate of the
polarization energy from Ceperley-Alder.\cite{cep80}  In that
work, size effects were estimated with Fermi liquid corrections.
Rather than BF-3B wavefunctions, corrections using the potentially
exact, release-node method were used.  The results with PBC and
FN-DMC are in agreement with the present FP-DMC calculations.
However, the present results have an error bar more than an order
of magnitude smaller than those of Ceperley-Alder,  primarily due
to increased computer performance.

Though PBC with FLT corrections are adequate for unpolarized and
fully polarized systems, the precision is limited for intermediate
polarizations. To estimate finite size effects within FLT, one
must perform accurate DMC simulations for widely varying system
sizes. In Ortiz et al.,\cite{ortiz} the simulation size varied
from $725 \leq N \leq 1450$. Within DMC it is very time-consuming
to ensure uniform accuracy independent of particle number, so that
one typically determines size effects within VMC, using the more
approximate SJ trial functions. As we have seen in Fig. (1), the
SJ trial functions are unreliable at low densities. TA boundary
conditions allow a much better way to estimate energies in the
thermodynamic limit of partially polarized fermi liquids since the
number of electrons can be held fixed as the spin polarization
varies. Small system sizes, allowing use of more accurate but
expensive trial functions and even exact fermion methods, give
precise estimates of the spin polarization energy in the
thermodynamic limit.

\begin{figure}
\centerline{\psfig{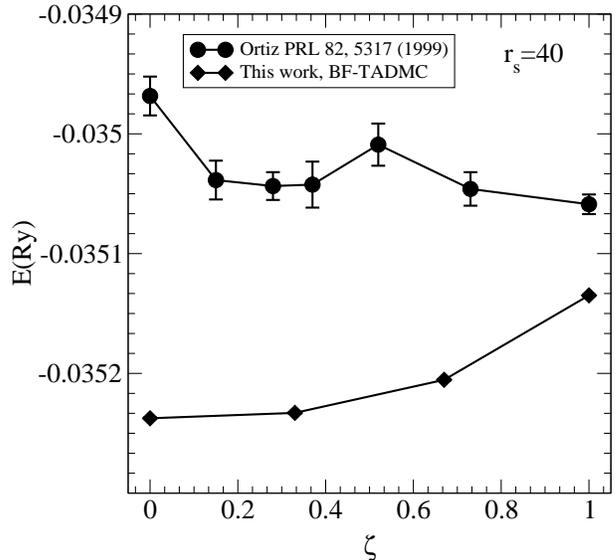}}
\caption{\label{TA_ortiz}Comparison of this work with that of
Ortiz et al.\cite{ortiz} The filled diamonds are DMC simulations
with TA and BF-3B wavefunctions (this work). The filled circles
are DMC with PBC and SJ wavefunction.\cite{ortiz} All energies are
extrpolated to the thermodynamic limit. Errors are given in Table
I and are smaller than the size of the points. }
\end{figure}

In Fig. \ref{TA_ortiz} we compare the total energy for $r_s=40$
calculated with DMC and TA and extrapolated to the thermodynamic
limit with the calculation of much larger systems ($N=725$) of
Ortiz et al. \cite{ortiz} The BF-3B energies are lower energy and
show a different polarization energy: Ortiz's calculation finds
that the polarized or partially polarized phase is stable at this
density, while we find the unpolarized phase is stable. This
difference is due to the backflow correlations in the trial
wavefunction. Although backflow energies are small, they favor the
unpolarized state and hence are crucial for accurate determination
of the polarization transition.

\section{Results}

We carried out computations of the spin polarization energies at
electronic densities $ 40 \leq r_s \leq 100$. At each density, we
performed DMC calculations with \( N=54 \) and \( N=108 \)
electrons using \( 10^{3} \) twist angles. The time step was
adjusted so that the DMC acceptance ratio was in the range \(
98\%-99\% \). Note that when calculating the polarization energy,
most time step errors will cancel out of the polarization energy.
Thus systematic errors in the polarization energy are much smaller
than in the total energy. We then extrapolated the energy to the
thermodynamic limit using Eq. (\ref{3deg-TA}). The energies are
given in Table I.
\begin{figure}
\centerline{\psfig{figure=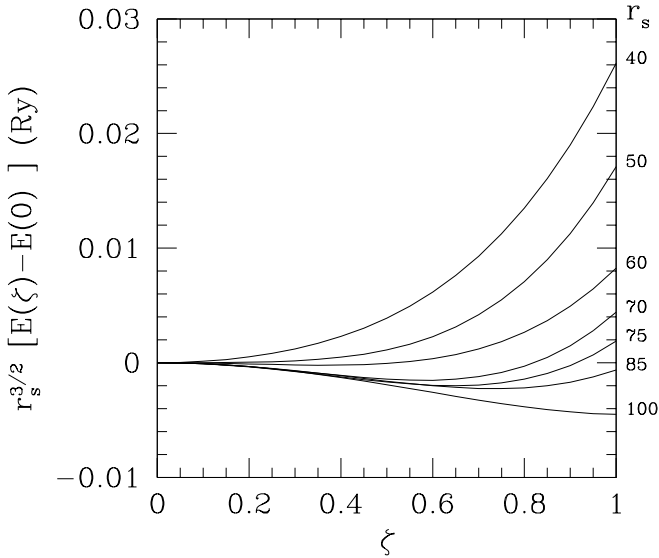,width=10cm,angle=0}}
\caption{\label{allfits}The spin polarization energy of the 3DEG
times $r_s^{3/2}$ in Ry/electron at various densities using a
polynomial fit to the data in Table I. The density, $r_s$, is
denoted on the right axis.}
\end{figure}
We then fit the energy versus polarization to a quadratic
polynomial in $\zeta^2$. The results are shown in figure
\ref{allfits}. A polarization transition is evident. At \(
r_{s}=40, \) the system is still paramagnetic, with the
unpolarized phase stable. As the density decreases, at \( r_{s}
\approx 50, \) the system becomes unstable with respect to spin
fluctuations.  The partially polarized states become stable at \(
r_{s}\geq 60. \) As the electronic density continues to decrease,
the fully polarized state has a lower energy with respect to
unpolarized state at \( r_{s} \geq 80 \), however, we find that
the partially polarized state has an even lower energy.

\begin{figure}
\centerline{\psfig{figure=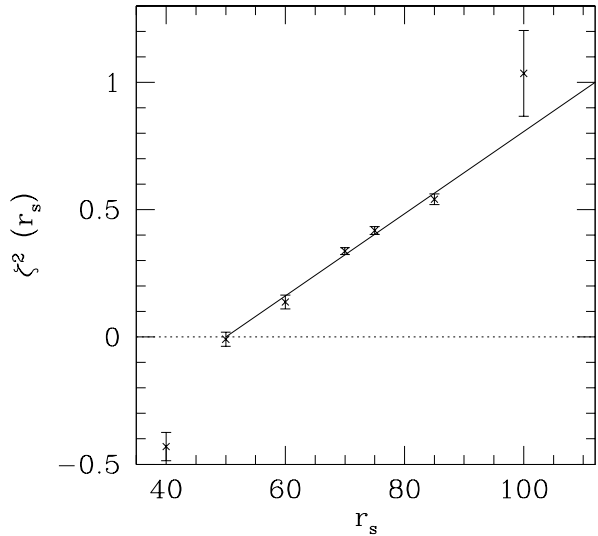,width=10cm,angle=0}}
\caption{\label{minpol}The square of the spin polarization versus
$r_s$. The curves were obtained using fits in the Fig. (4). The
line is a fit though the points. The value at $r_s=40$ was
obtained by extrapolation from physical values of $\zeta$. }
\end{figure}

In Fig. \ref{minpol} is shown the predicted square of the optimal
polarization versus density. We find that the equilibrium
polarization is described by $\zeta^2 = (r_s-r_s^*)/62$ with the
critical density $r_s^*= 50 \pm 2$. As the density decreases, the
stable state becomes more and more polarized approach fully
polarized at freezing, $r_s \approx 100$. Quantum critical
fluctuations, not present in systems with $N \leq 162$, could
modify the behavior of the spin polarization energy near the
critical density.

The quoted error bar on the critical density estimates the
statistical errors, not the systematic errors arising from the
fixed-phase approximation.  The experimental and theoretical
results on polarized helium give caution on placing too much
confidence in the estimate of the polarization transition. Even
using the accurate optimized BF-3B wavefunctions, the magnetic
susceptibility in liquid $^3$He does not agree with experiment at
low pressure and the polarized phase is nearly degenerate with the
unpolarized phase at the freezing density.\cite{moroni} The
present results also do not preclude the existence of phases with
other order parameters such as superfluids, as occurs in ground
state of liquid $^3$He. In fact, it is rather likely that the
ground state of the electron gas will have such a phase at the
lowest fluid densities.

\begin{figure}
\centerline{\psfig{figure=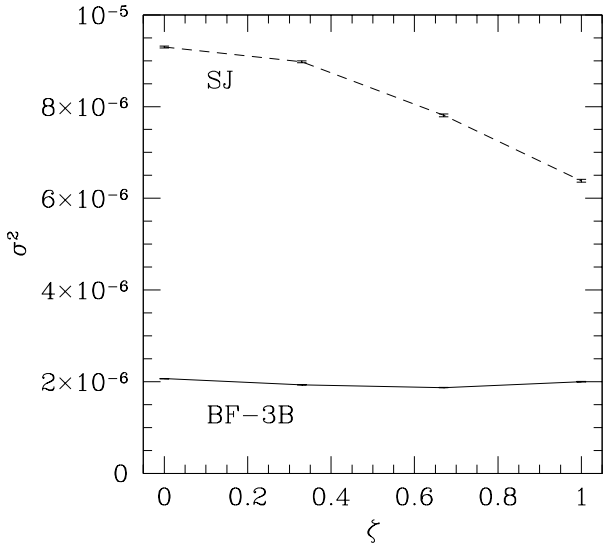,width=10cm,angle=0}}
\caption{\label{variance}The variance of the SJ and BF-3B trial
function as a function of spin polarization at \protect{$r_s=50$}
using TA and VMC with $N=54$. }
\end{figure}

However, examination of the variance of the trial function
suggests that the result for the electron gas may be more reliable
than for liquid $^3$He. Shown in fig. \ref{variance} is the
variance of the trial function at $r_s=50$ for both the SF and
BF-3B functions. Although the variance of the SJ trial function
depends on spin polarization, that of the BF-3B does not. Such is
not the case\cite{moroni} with liquid $^3$He. Arguments based on
variance extrapolation\cite{kwon93} suggest that the DMC
calculations with BF-3B phases should be more reliable than in
liquid $^3$He.

\section{The Phase Diagram}

One can use the calculated energies to estimate the finite
temperature behavior within the Stoner model,\cite{stoner} arising
in the theory of itinerant magnetism.\cite{herring} The Stoner
model differs from Hartree-Fock by replacing the Coulomb
interaction by a zero range one, a repulsive delta function
potential: $\sum_{i<j} g \delta(r_{ij})$. One can view this
approach as the first step to a full Fermi-liquid description of
the quasiparticle interactions, and use the QMC data to determine
the strength of those interactions. One expects that the
Slater-Jastrow trial function has screened off the long-range
interaction leaving only a short-ranged spin-dependent term that
can be modeled by a contact interaction.

In the Stoner model, the energy is evaluated within the mean field
(Hartree-Fock) approximation using the NI wavefunction. The energy
at zero temperature in the thermodynamic limit is: \be E \propto
(1 + \zeta )^{5/3} +(1-\zeta)^{5/3} + 0.054 g r_s^2 (1 - \zeta^2).
\ee  For $gr_s^2< 20.5$ the system has an unpolarized ground state
and for $gr_s^2>24.4$ the ground state is ferromagnetic. For
intermediate couplings, the ground state has a partial spin
polarization at zero temperature, similar to the observed behavior
of the electron gas at low density.

Although the polarizations are qualitatively correct, the above
functional form does not fit well the DMC data ({\it i.e.} from
Table I.) In addition, assuming $g$ does not have a very strong
density dependence, the Stoner model predicts that the partially
polarized density range should be quite narrow, from $50 \le r_s
\le 54$, while as the QMC results indicate a much broader density
range. Perhaps the assumption of a zero-range interaction of
quasiparticles is too restrictive.

\begin{figure}
\centerline{\psfig{figure=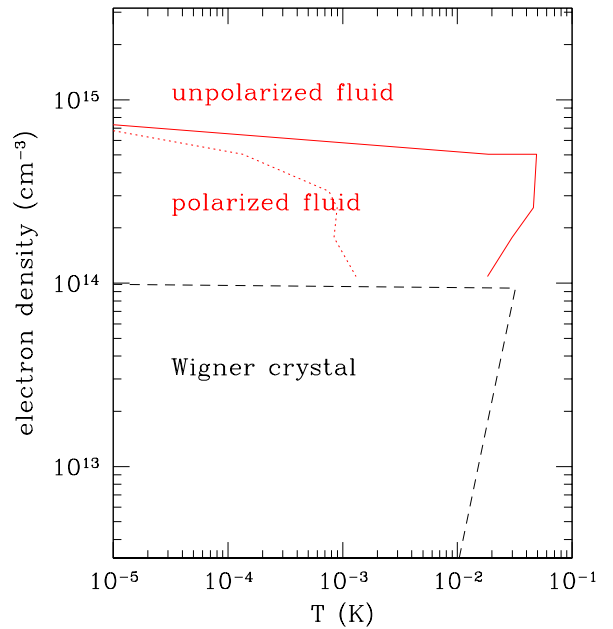,width=10cm,angle=0}}
\caption{\label{egpd}The phase diagram of the electron gas.
Conversion to units of cm and K was done using $a_0=1.3 nm$ and
$Ry=250K$ using estimates\cite{pickett} of the effective mass and
the dielectric constant of $SrB_6$. The solid line the mean-field
estimate of the magnetic transition temperature from the Stoner
model, where the spin interaction is estimated from the zero
temperature QMC data. The dotted line is the energy difference
between the unpolarized and partially polarized system.
}\end{figure}

Nevertheless, we use this mean field model to make a estimate of
the transition temperature of the polarized phase.  Fitting the
energy to the function given in Eq. (8), we obtain estimates of
$g$ and the effective mass of the electrons. Then, using the free
energy of the ideal spin (1/2) partially polarized fermi gas with
the additional spin interaction, we estimate the temperature at
which the system becomes polarized by determining when the
spin-stiffness of the unpolarized system vanishes. Fig. \ref{egpd}
is the estimated phase diagram of the electron gas.\cite{suris} In
this diagram, the effective mass and dielectric constant for
$SrB_6$, a closely related material to $CaB_6$, have been
used\cite{pickett} to convert to units of $K$ and $cm^3$. Note
that both the temperatures and densities of our calculated
magnetic transition are four orders of magnitude smaller than that
found experimentally\cite{young99} in $CaB_6$. Even assuming
errors because of uncertainties in material properties and from
the mean field methods to estimate $T_c$, these estimates are very
difficult to reconcile with experiment, apparently ruling out an
electron gas model of ferromagnetism in this material. Also
plotted on the phase diagram is the energy difference between the
partially polarized fluid and the unpolarized fluid as another
estimate of the magnetic transition temperature. Finite
temperature QMC calculations would be desirable to confirm the
mean field estimate of $T_c$. A rough estimate of the limit of
stability of the Wigner crystal \cite{jones96} is also shown.

Tanaka and Ichimaru\cite{ichimaru} have computed the polarization
phase diagram of the electron gas both at zero and non-zero
temperature using an integral equation method.  At zero
temperature they obtain a result similar to Ortiz at
al.\cite{ortiz}, with a continuous transition at $r_s \approx 20$
and a fully polarized fluid state at a slightly higher density
$r_s \approx 22$. Apparently this approach, built on local field
corrections to free fermion response functions, is biased by the
initial Hartree-Fock assumption and overemphasizes the tendency
for ferromagnetism.

There is recent work on the low density electron gas in 2D,
studying both the Fermi liquid\cite{atta} and the Wigner crystal
phase.  Using QMC similar techniques  in the liquid phase, a
polarized phase is found to be stable between $26 \leq r_s \le
35$, though the energy differences are even smaller than in 3D.
The partially polarized phase is never stable. In the 2D Wigner
crystal, path integral methods\cite{bernu} were used to derive
directly the spin Hamiltonian.  It was found that the ground
magnetic state is a spin liquid though the ferromagnetic state has
only a slightly higher energy at melting. Analogous calculations
of the magnetic phase diagram of the WC in 3D are underway.
\cite{candido}

\section{Conclusion}

We studied the polarization transition in three dimensional
electron gas using twist-averaged boundary conditions and trial
functions with backflow and three body correlations. Twist
averaged boundary conditions have a much reduced systematic
finite-size error, especially for the calculation of polarization
energies, enabling size-converged results with fewer electrons.
Using relatively small system sizes allows one to use more
accurate trial wavefunctions and to fully converge the diffusion
Monte Carlo calculations. We find a second order transition to a
polarized phase at $r_s = 50 \pm 2$.

In general, methods based on the variational principle such as the
fixed-node quantum Monte Carlo method for many-fermion systems,
favor phases with a higher symmetry, in this case the Wigner
crystal and polarized fluid, over the more complicated unpolarized
phase, (effectively a two-component mixture of spin up and down
electrons.) Recall that in HF one has a polarization transition
because antisymmetry is the only way to correlate electrons,
however, the correlation is only between like spins; hence, there
is an instability to polarize the system once the potential is
dominant. But using a SJ wavefunction both like and unlike
electrons are highly correlated. Our results demonstrate that the
SJ wavefunctions still preferentially favors the ``simple''
phases, even using DMC method.  This symmetry argument explains
the historical tendency for the polarized phase become stable over
a more and more restricted range of density and temperature as
more accurate methods are used.  Our QMC results using the BF-3B
wavefunction indicate that there is still an instability for spin
polarization at very low density.  Although examination of the
variance indicates that the polarized and unpolarized BF-3B trial
functions are equally inaccurate, it is still not clear if our
finding of a polarization transition is an artifact of the assumed
trial wavefunction. Calculations with the new methods (TA and
BF-3B wavefunctions) but with the exact fermion methods are
desirable to resolve the phase of the electron gas at intermediate
densities.

\acknowledgements {This research was supported by NSF DMR01-04399,
the Department of Physics at the University of Illinois
Urbana-Champaign and the CNRS. We acknowledge unpublished data
from G. Ortiz shown in fig. (3). Computational resources were
provided by the NCSA.}

\end{multicols}
\widetext \vspace*{-5mm}

\begin{table}\label{TableI}
\caption{Energy of the 3DEG computed using TA- DMC and
extrapolated from \( N=54 \) and \( N=108 \) with (\( 10^{3} \)
twist angles). Energies are in Ry/electron. The numbers in
parenthesis are standard errors in units of $10^{-8} Ry$. }
{\centering
\begin{tabular}{cccccccc}
$r_s  \backslash \zeta$&
0.0&0.185&0.333&0.519&0.667&0.852&1.0\\
\hline  40 & -0.03523748(60)& & -0.03523295(67)& &-0.03520539(67)& &-0.03513483(72)\\
50 & -0.02889900(62)& -0.02889900(66) &-0.02889962(68)&
-0.02889449(70)&-0.02888835(62)& -0.02887542(70) &-0.02884983(81)\\
60 & -0.02452017(44)& -0.02451866(51)& -0.02452031(48)&
-0.02451963(50)& -0.02451747(42)& -0.02451188(46)&
-0.02450167(46)\\
70 & -0.02131429(41)& -0.02131381(40)& -0.02131621(39)&
-0.02131716(37)& -0.02131593(37)& -0.02131332(39)&
-0.02130667(37)\\
75 & -0.02001137(35)& -0.02001191(37)& -0.02001376(36)& &
-0.02001434(44)& &
-0.02000878(33)\\
85 & -0.01784017(30)& & -0.01784152(32)& & -0.01784300(32)& &
-0.01784109(32)\\
100 & -0.01535357(30) &  & -0.01535340(30) &  & -0.01535639(26)&
 & -0.01535761(26)\\
\end{tabular}
\par}
\end{table}
\begin{multicols}{2}

\end{multicols}

\begin{thebibliography}{10}
\bibitem{ortizm}The method in ref. (1) is essentially the same
as in ref. (6), though there are several technical differences:
more particles, variance minimization of short-ranged wavefunction
rather than an analytic form with correct long-range correlations,
and the use of a fixed-node method with a Slater Determinant of
plane waves.
\end{thebibliography}
\end{document}